\documentclass[%
 reprint,
 superscriptaddress,
 amsmath,amssymb,
 aps,
]{revtex4-2}

\usepackage{graphicx}
\usepackage{dcolumn}
\usepackage{bm}
\usepackage{xcolor}

\newcommand{\KK}[1]{\textcolor{black}{\,#1}}

\begin{document}


\title{Model-free stiffness measurement of microscale soft contacts by far-field traction force microscopy}

\author{Keumkyung Kuk}
\affiliation{Laboratory for Soft Materials and Interfaces, Department of Materials, ETH Zürich, 8093 Zürich, Switzerland}
\author{Shivaprakash N. Ramakrishna}
\affiliation{Laboratory for Soft Materials and Interfaces, Department of Materials, ETH Zürich, 8093 Zürich, Switzerland}
\author{Nancy A. Burnham}
\affiliation{Empa, Swiss Federal Laboratories for Materials Science and Technology, Dübendorf, 8600, Switzerland}
\affiliation{Department of Physics, Worcester Polytechnic Institute, Worcester, MA, 01609, United States}
\author{Robert W. Style}
\affiliation{Laboratory for Soft Materials and Interfaces, Department of Materials, ETH Zürich, 8093 Zürich, Switzerland}
\author{Lucio Isa}
\affiliation{Laboratory for Soft Materials and Interfaces, Department of Materials, ETH Zürich, 8093 Zürich, Switzerland}

\date{\today}

\begin{abstract}
The microscale stiffness of soft materials governs their contact mechanics, adhesion and friction, yet it is surprisingly difficult to measure. 
Micro or nanoindentation is typically used, but extracting stiffness from these measurements requires using contact mechanics models (e.g. JKR), whose assumptions often break down for soft adhesive contacts.
Here, we introduce far-field traction force microscopy (FF-TFM) to extract stiffness without using any contact model. 
We drag a colloidal probe mounted on an atomic-force microscope across a soft surface, measuring the friction force while imaging the displacements of fluorescent tracers on the substrate. 
Far from the contact, in the elastic limit, these lateral displacements depend solely on the friction force and the substrate stiffness, and fitting them with the point-force solution thus gives the substrate's Young's modulus directly. 
We successfully match bulk rheology moduli with FF-TFM for six different silicone elastomers, where JKR fits to indentation data underestimate moduli by as much as 40\% of their bulk values. 
Our approach also enables tracking how the probe slips across the surface. 
We find that soft contacts can `pre-slip’ with no detectable signature in the force signal, unlike stiffer contacts.
FF-TFM thus offers a contact-model-free route to measuring microscale stiffness and slip behavior in soft materials.
\end{abstract}


\maketitle

\section{Introduction}

Stiffness is a key material property. It controls how much a material deforms under an applied force and its knowledge is thus essential to design any structural or mechanical component, in particular whenever contacts between two (or more) bodies are present. 
While the mechanistic origin of contact stiffness has been exhaustively investigated for hard surfaces, i.e. between two metals, the advent of a broad emerging class of new soft materials, such as functional elastomers or hydrogels, requires new methodologies and new understanding.
Measuring stiffness of soft materials at the macroscale is now a routine task. It is carried out either by tensile testing to obtain a material's Young's modulus ($E$) or via shear rheology to measure its shear modulus ($G$), whereby $E$ can be obtained through knowledge of the material's Poisson's ratio $\nu$.  

However, it is extremely challenging to transfer these tests onto microscopic samples, and this limitation presents far-reaching implications. Microscale contacts in fact govern the performance of many soft materials of high relevance, where surface deformation controls not only the contact area, but also adhesion and friction~\cite{Schallamach1971HOWSLIDE,Barquins1985SlidingReview,Viswanathan2022FiftyFracture,Djellouli2026SqueakingInterfaces}. 
In particular, living systems are highly sensitive to microscale contact stiffness, where cells sense the stiffness of their immediate surroundings and change their behaviour in response~\cite{Wells2008TheBehavior,Guimaraes2020TheEngineering}.
For example, macrophages sense the stiffness of an implant surface, and the response determines how much inflammation and scarring will follow~\cite{Ni2023MacrophagesDeformation}.
Matching the contact stiffness of an implant with its host tissue has thus become a key goal in bioelectronics design~\cite{Oh2024SofteningDirections}.
Contact stiffness also matters for most soft devices.
For example, the sensitivity of soft tactile sensors is set by the stiffness of structures only a few hundred micrometres in size~\cite{Yun2026Multiscale-structuredSensors}.
Soft robotic grippers also often rely on micropatterned sensors and small scale contacts under conditions of varying friction and adhesion, e.g. wet versus dry contacts~\cite{Yao2025FlexibleApplications,Adachi2026AI-integratedEnvironments,Li2026MicroscaleSystems}.
In all of these cases, it is vital to characterize and engineer microscale contact stiffness.

Nonetheless, substantial difficulties prevent the accurate measurement of stiffness at the microscale.
Small-scale samples cannot be cut out and clamped to standard measuring tools, they are frequently too small or thin for bulk rheometry, and too stiff ($E\gtrsim 100$\,kPa) for alternative techniques such as microrheology~\cite{Liamas2020ProbingNanoscale,Joyner2020MicrorheologyDesign}. 
As a result, one typically relies on indentation-based techniques~\cite{He2024AMaterials,Norman2021MeasuringMicroscopy}.
Unfortunately, indentation does not measure stiffness directly.
It measures the force and displacement of a probe moving against a substrate and uses a contact mechanics model to extract a modulus from their fitting. 
The most common models, such as the Hertz or JKR (Johnson-Kendall-Roberts,~\cite{Johnson1971SurfaceSolids}) models, come with assumptions of small strains, homogeneous elastic half-space, and adhesion that is either absent (Hertz) or constant (JKR), which often break for soft microscale contacts~\cite{Style2013SurfaceSolids,Petroli2022DeterminationMicroscope,Onozuka2024AtomicContact,He2026NonmonotonicHydrogels}.

Here, we demonstrate an alternative approach for measuring the stiffness of soft substrates at the micro-scale that requires no fitting to a contact model. 
We indent substrates with a colloidal probe mounted on an Atomic Force Microscope (AFM) and drag the probe laterally while measuring the friction force required to do so. 
Simultaneously, we measure the displacements of the free substrate surface around the probe position. 
Far from the probe, these displacements depend only on the applied force and the substrate stiffness, and not on what happens within the contact, which is precisely what contact models attempt to describe.
Combining force measurements and far-field substrate deformations thus allows us to extract substrate stiffness without needing to know details about the probe/substrate contact area.
We apply our approach to silicone elastomers, demonstrating that this technique outperforms classical indentation approaches, with extracted values of $E$ closely matching those obtained from bulk rheology on macroscopic samples. 

Beyond measuring stiffness, we also show that our technique  allows the precise measurement of slip during friction experiments.
On stiff materials, slip before full sliding shows up as a small change in the friction force~\cite{Rubinstein2007DynamicsSliding,Katano2014NovelSlipping,Maegawa2010PrecursorsLoading}.
On soft materials, however, we find that the manifestation of slip can be ``silent", i.e.
the probe slips across the surface with no detectable force signals. As a consequence, slip makes the contact look softer than it really is.
Unravelling ``silent" slip is particularly relevant wherever soft materials are used to grip, from biological tissues to soft robotics.

\section{Results}
We showcase the difficulties in extracting stiffness from small-scale indentation measurements by performing experiments on two different types of silicone elastomers. 
The first one is a standard commercial silicone (Sylgard 184, Dow), prepared by combining base and curing agent together with a mass ratio of 35:1 (labelled as S35:1). We term this `as-prepared'. The second silicone is made from the same material, but we use toluene to extract uncrosslinked oligomers after curing (see Materials \& Methods). 
We term this `solvent-extracted'. 
Both substrates are soft and of similar stiffness with Young's moduli of the order of 100 kPa measured via bulk rheology (Fig \ref{fig1}).
However, as we will see below, they exhibit strong differences in friction experiments.
The as-prepared silicone has a slippery surface, which we attribute to its significant uncrosslinked oligomer -- viscous liquid -- content (sol fraction: 28 wt.\%).
By contrast, the solvent-extracted silicone sticks strongly to the probe.

\subsection{Normal indentation}
\begin{figure} 
    \centering
    \includegraphics[width=1\linewidth]{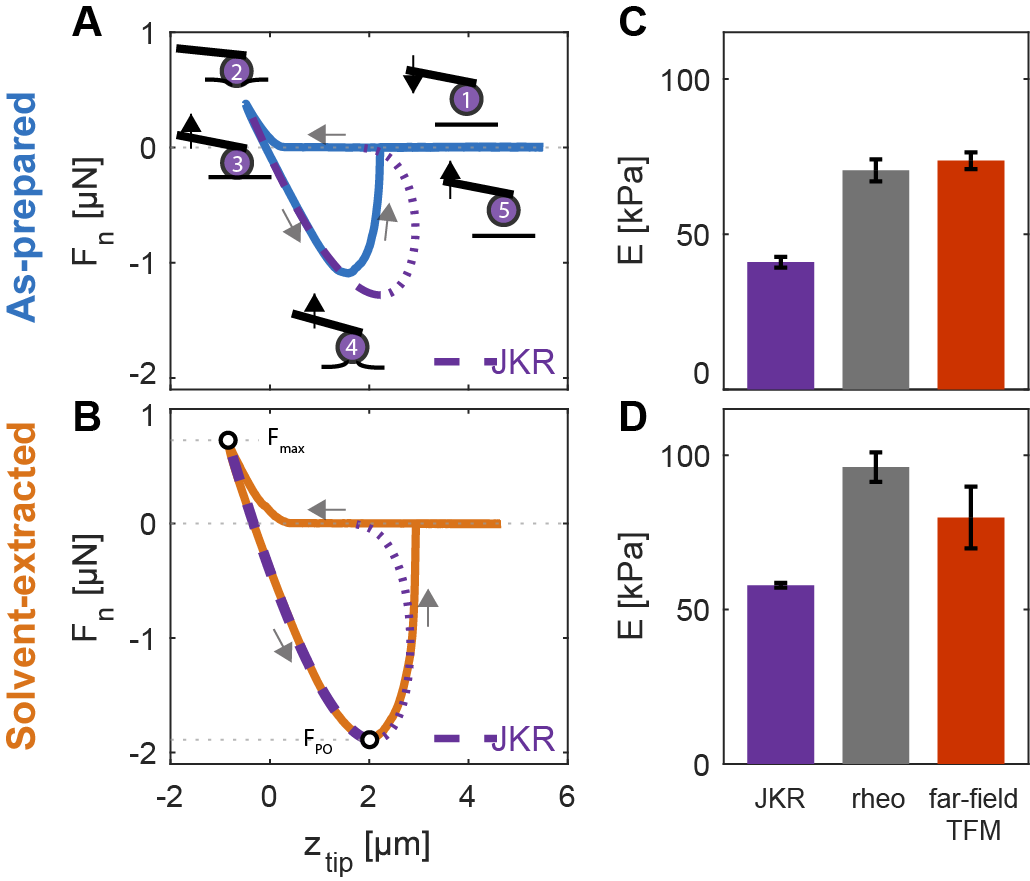}
    \caption{Extracting stiffness from contact measurements on as-prepared (top row) and solvent-extracted (bottom row) S35:1 silicone. A,B) Normal force, $F_n$, versus tip height, $z_{tip}$, during indentation and retraction (as indicated with arrows). Dashed curves show JKR fits to the retraction branch between $F_{max}$ and $F_{PO}$, and dotted lines show the fits extrapolated to larger $z_{tip}$. The schematics illustrate the approach, indentation, retraction and pull-off. 
    C,D) Young's modulus from JKR fits, bulk rheology and far-field TFM. Error bars show the standard deviation over 25 curves (JKR), $\pm5\%$ (rheology), and the standard deviation over the friction loop (far-field TFM).}
    \label{fig1}
\end{figure}

We first try to measure the stiffnesses of the two substrates using classical micro-indentation experiments.
We use an AFM cantilever with an attached spherical probe to slowly perform an indentation cycle on each substrate, immersed in water.
This measurement gives us the normal force, $F_n$, as a function of the height of the probe tip, $z_{tip}$.
Here, the probe is a smooth glass sphere of radius $R=40\,\mu$m (surface roughness: 1.3\,nm RMS, \KK{Fig S1}), which we move at an indentation velocity $\dot{z}_{tip}=0.5\,\mu$m/s.
The resulting raw force curves are shown in Fig \ref{fig1}A,B for the as-prepared and solvent-extracted samples, respectively. 
In both experiments, we see a similar overall behaviour. As the sphere is pushed into the substrate, the force increases up to a target normal force, $F_{max}$. 
Next, the sphere is retracted. 
$F_n$ reaches a maximum tensile (negative) `pull-off force' ($F_{PO}$) before it detaches from the substrate, and $F_n$ returns to 0. 
In both cases, $F_{PO}\gg F_{max}$, indicating that there is significant adhesion between the substrate and the probe -- as is generally found in micro-indentation experiments~\cite{Style2013SurfaceSolids}.

We extract stiffnesses from these datasets using JKR (Johnson-Kendall-Roberts) theory. 
This is a contact model that includes both elasticity ($E$) and adhesion (the work of adhesion, $W$), and which is applicable for small indentations~\cite{Johnson1971SurfaceSolids}.
(Note that an equivalent solution was found earlier by Sperling, in his unpublished doctoral thesis~\cite{Sperling1964EineKorpern,Li2020Non-adhesiveArea}.)
The JKR model comprises two equations~\cite{Shull2002ContactSolids}.
The first one relates indentation, $\delta$, to the contact radius, $a$:

\begin{equation}
    \delta=\frac{a^2}{3R}+\frac{F_n (1-\nu^2)}{2Ea}
    \label{eqn:JKR1}
\end{equation}

$\delta=z_0-z_{tip}$ measures how far the probe has indented below the position of the undeformed surface ($z_0$).
Here, $\nu$ is the substrate's Poisson ratio (we take $\nu=1/2$, as this is true for most soft materials~\cite{Pritchard2013PreciseCorrelation}), and we have assumed that the probe's material is much stiffer than the substrate. 
The second JKR equation relates $a$ to the substrate `stickiness', i.e. $W$:

\begin{multline}
    a^3=\frac{3R(1-\nu^2)}{4E}\biggl(F_n+3\pi W R \\
    +\sqrt{6\pi W R F_n+(3 \pi W R)^2} \biggr).
    \label{eqn:JKR2}
\end{multline}

To apply these equations to our model, we must assume that $W$ is constant during the retraction process.
Then, we can combine the two equations to eliminate $a$, yielding an equation for $\delta(F_n,E,W,\nu,R)$~\cite{Grierson2005AccountingMicroscopy,Ebenstein2011Nano-JKRWater}.
We finally match this expression to the retraction phase of the curve $F_n(z_{tip})$, by fitting the unknown parameters $E$, $W$, and $z_0$ via a nonlinear-least-squares fitting~\cite{Ebenstein2011Nano-JKRWater}.
For the two different experiments, the fits to the data in Fig \ref{fig1}A,B are shown as dashed/dotted curves.
The dashed parts of the curves show the fitted range of $z_{tip}$ up to the pull-off. 
The dotted parts of the curves show the fitted theory extrapolated to larger $z_{tip}$.
For the as-prepared substrate, we fit $E=41.0$ kPa and $W=6.7\,\mathrm{mJ/m^2}$.
For the solvent-extracted substrate, $E=57.9$ kPa and $W=10.2\,\mathrm{mJ/m^2}$.

JKR theory appears to do an excellent job of fitting the retraction phase of the data.
However, we see issues when we try to extrapolate the theory to describe unfitted parts of the indentation curves, or when we compare fitted values of $E$ to values measured from bulk experiments.
For example, in the as-prepared silicone, the extrapolated (dotted) theory does a poor job of matching the data -- significantly over-predicting the pull-off force.
Furthermore, values of $E$ from the JKR fit are only around 60\% of the values that we measure via bulk rheology (see Materials \& Methods), as shown in Fig \ref{fig1}C.
This is even true for the solvent-extracted sample (Fig \ref{fig1}D), which appears to be extremely-well fitted by JKR theory.

These observations suggest that fitting indentation experiments with JKR theory may not lead to accurate measurements of $E$.
In fact, this is unsurprising.
Experimentally, there are constraints on the accuracy of the technique that can bias force measurements~\cite{Pratt2010QuantitativeMicroscopy} (see \KK{SI, section S1}).
Further, in the JKR theory, there are several underlying assumptions that can commonly break down.
For example, the fitting assumes that $W$ is constant during the retraction process.
However, this is certainly not true: previous experiments have shown how $W$ can increase by orders of magnitude in this period~\cite{Shull2002ContactSolids}, due to dissipative processes at the substrate/probe contact line, or adhesion hysteresis caused by surface roughness~\cite{Onozuka2024AtomicContact,He2026NonmonotonicHydrogels,Dalvi2019LinkingRoughness,Sanner2024WhyThem}.
Furthermore, JKR theory assumes small strains everywhere.
However, this is not true in a very localized region near the contact line on a soft substrate, where the substrate is stretched into a sharp neck (see schematic in Fig \ref{fig1}A)~\cite{Style2013SurfaceSolids}.
Finally, in the as-prepared silicone, we have to worry about the presence of uncrosslinked oligomers.
These free oligomers are known to alter the behaviour of probe/substrate contact lines, with oligomers being locally extracted from the substrate, and wetting the probe surface~\cite{Jensen2015WettingAdhesion,Wong2020AdaptivePolydimethylsiloxane,Glover2023CreasingFriction}.
With all of these potential sources for error, it is thus not surprising that we can have apparently excellent curve fits, while obtaining erroneous values of $E$.

\subsection{Far-field TFM: traction force microscopy far away from contact area}
\begin{figure*}[t!]
    \centering
    \includegraphics[width=1\linewidth]{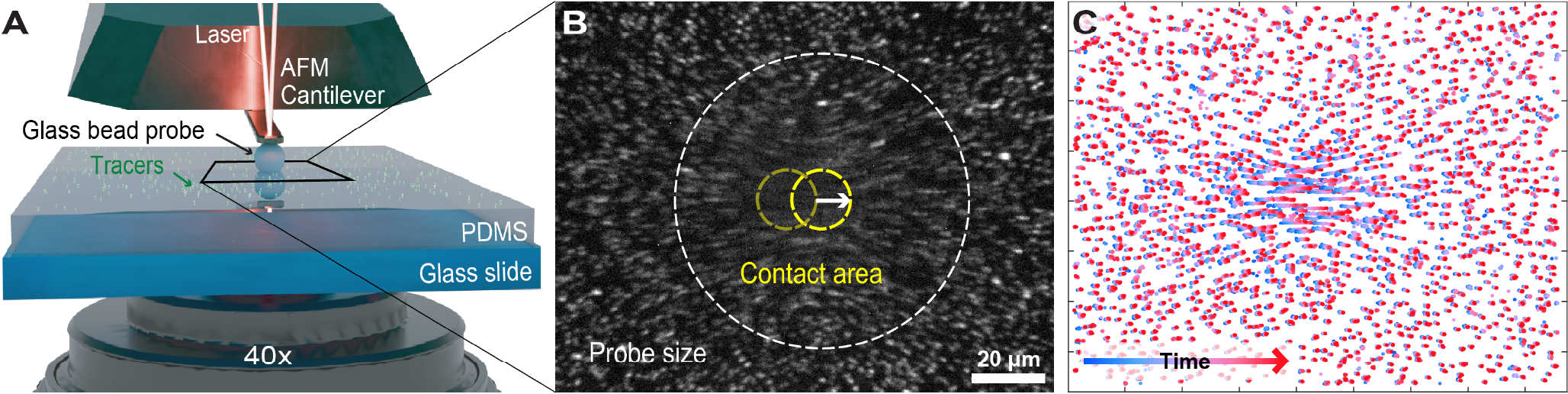}
    \caption{Performing an FF-TFM experiment. A) Schematic of the setup. A spherical probe on an AFM cantilever is indented into, and then pulled laterally across the surface of a silicone substrate coated with fluorescent tracers. Tracers are imaged from beneath with a confocal microscope. B) A long-exposure image of tracers on a substrate surface (solvent-extracted S35:1) during a friction experiment (normal force $F_n=$ 350 nN, scanning velocity $\dot{x}_{tip}=0.5\,\mu$m/s) shows how the surface is pulled laterally near the probe/substrate contact area. Yellow circles show the movement of the contact area between the start and end of the experiment. The white dashed circle marks the size of the probe. Scale bar: $20\,\mu$m. C) Tracked tracer positions over the same frames. Each position is coloured by time, from blue (start of lateral loading) to red (near maximum lateral force).}
    \label{fig2}
\end{figure*}

The fact that JKR theory appears to give unreliable measurements of $E$ raises the question of whether it is possible to measure stiffness via micro-indentation experiments more accurately and without assumptions on the details of the probe/substrate contact area.
In particular, we propose to sidestep those assumptions by relating applied forces to substrate displacements measured around the outside of a substrate-probe contact.
We call this technique `Far-Field Traction Force Microscopy' (FF-TFM), due to its close conceptual ties to Traction Force Microscopy~\cite{Style2014TractionBiology}.
We perform sliding friction experiments, where we indent an AFM-mounted spherical probe into the substrate, and then slowly pull it parallel to the surface (Fig \ref{fig2}A).
We record the lateral force, $F_{lat}$, by measuring the torsion of the AFM cantilever (see Materials \& Methods) and, simultaneously, we measure displacements of the substrate surface under and around the contact patch via confocal microscopy from beneath (i.e. through the transparent silicone).
We do this by attaching a dilute number of fluorescent nanoparticles (`tracers') to the substrate surface, and imaging their displacements as the experiment progresses (Fig \ref{fig2}B,C, details in Materials \& Methods).

An example of a typical tracer displacement map is shown in Fig \ref{fig2}B,C.
Here, we push the probe into a solvent-extracted substrate (S35:1) with a normal force of 350\,nN, and then slowly pull it laterally (in the $x$-direction) at scanning velocity, $\dot{x}_{tip}=0.5\,\mu$m/s over a distance of $\sim5\,\mu$m.
The whole experiment is submerged in water.
Figure \ref{fig2}B shows a long exposure, fluorescence image of the substrate surface during this movement.
Tracers (bright spots) near the probe/substrate contact area are dragged laterally, while tracers further away do not significantly move, see \KK{Movie S1} for the corresponding time-lapse video.
We analyse these data by tracking the position of tracers at time points during the lateral motion (see Materials \& Methods).
Figure \ref{fig2}C shows how these positions evolve with time, with blue points showing the zero time (`force-free') state after normal indentation, but before lateral sliding occurs.

\begin{figure*}[t!]
    \centering
    \includegraphics[width=0.65\linewidth]{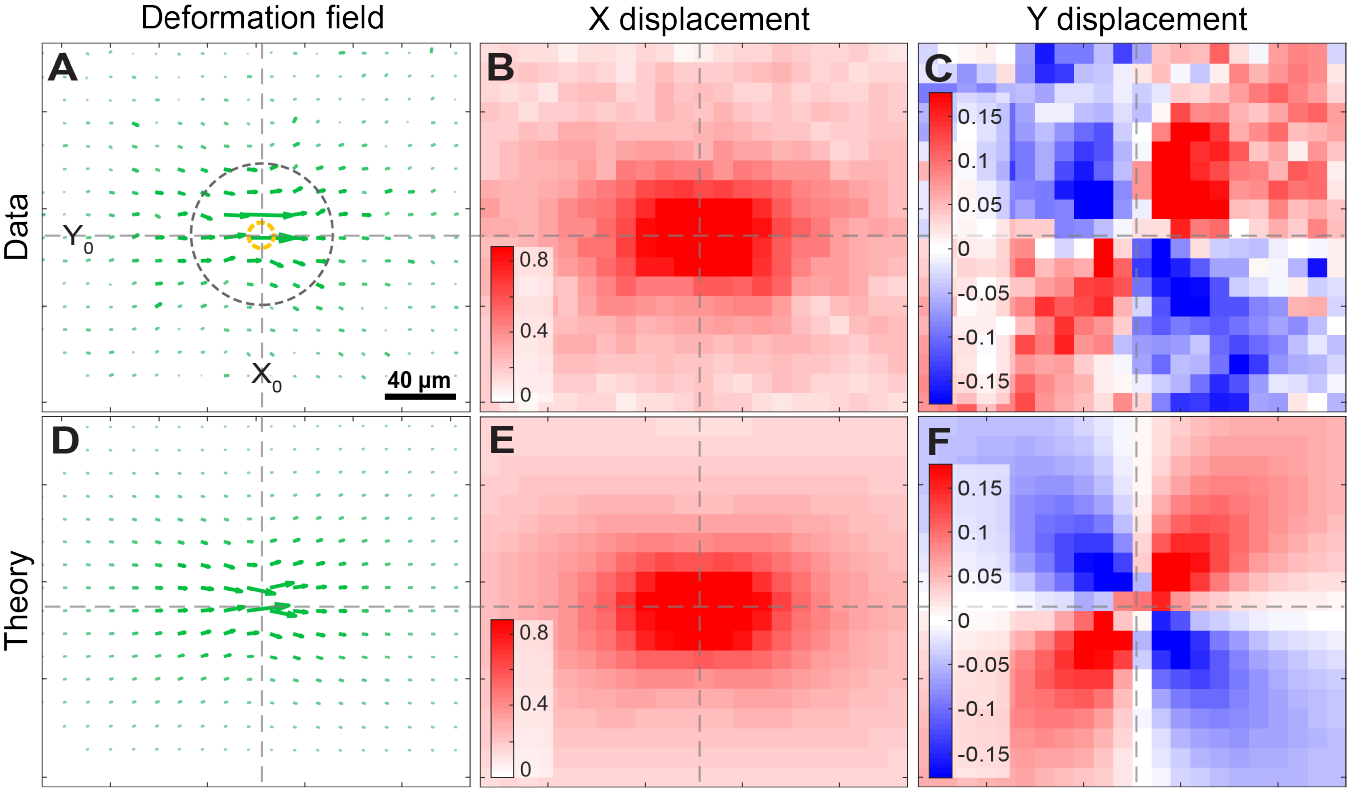}
    \caption{Surface displacements as a probe slides across the surface of a substrate (Data, upper row), and the point-force fit (Theory, bottom row). A) Measured displacement field for the data in Fig \ref{fig2}B,C, interpolated onto a grid, scale bar: $40\,\mu$m. The grey dashed circle marks the size of the probe, and the yellow dashed circle marks the contact area. B,C) The $x$ and $y$ components of the measured displacements, $u_x$ and $u_y$. D-F) The same quantities for the fitted point-force solution, equations (\ref{eq:bouss_u_Rob},\ref{eq:bouss_v_Rob}). Dashed lines mark the fitted force origin, $(X_0,Y_0)$. The unit of the colour scale is $\mu$m.}
    \label{fig3}
\end{figure*}

We further analyse the tracer motion by calculating tracer displacements, and interpolating the results onto a grid.
We define vector displacements $\mathbf{u}=(u_x,u_y)$, as the lateral movements of tracer particles relative to their positions in the force-free state:
Figure \ref{fig3}A shows $\mathbf{u}$ corresponding to the raw data in Fig \ref{fig2}C.
This plot is centred on the force-free position of the probe before movement. As it might be expected, we see large displacements near the contact, decaying to zero over a characteristic distance that is comparable to $a$.
Figure \ref{fig3}B,C plot the displacement components in the $x$ and $y$ directions, $u_x$ and $u_y$.
These are plotted as functions of the position on the surface in the force-free state ($X,Y$). 
The $x$ components highlight the large displacements following the probe movement: there is a strongly deformed, elliptical region of high displacements, with its long axis in the sliding direction.
By contrast, $y$ displacements are much smaller.
These form a characteristic checkerboard pattern, with alternating sign in the four quadrants surrounding the probe.
This pattern is a consequence of substrate stretching/compression around the probe.
Behind the probe, the substrate is stretched in the $x$ direction, causing it to contract vertically in the $y$ direction.
Ahead of the probe, the substrate is compressed in the $x$ direction, having the opposite effect.

To extract information about surface stiffness from these data, we note these fields are essentially identical to displacements that result from a lateral point force $(F_{lat},0,0)$ applied to the surface of a flat, incompressible ($\nu=1/2$), elastic substrate. This `Cerruti solution' is given by 
\begin{align}
  u_{x}^C &= \frac{3 F_{lat}}{4\pi E}
    \left(
      \frac{1}{r} + \frac{(X-X_0)^2}{r^3}
    \right) 
  \label{eq:bouss_u_Rob} \\
  u_{y}^C &= \frac{3 F_{lat}}{4\pi E}
    \left(
      \frac{(X-X_0)(Y-Y_0)}{r^3}
    \right),
  \label{eq:bouss_v_Rob}
\end{align}
where $r=\sqrt{(X-X_0)^2+(Y-Y_0)^2}$ is the radial distance from the centre of force application ($(X_0,Y_0)$ in the force-free state).
Figure \ref{fig3}D-F show the typical form of the resulting displacements and how they clearly match the measured data.
In fact, the agreement between the two is expected, following from St. Venant's principle. 
This principle states that the displacement fields from a point force and a stress distribution (i.e. caused by a probe) are indistinguishable far from the point of force loading -- provided that the total force, $F_{lat}$, is the same, and that the centre of the stress distribution is at the same position as the point force. Here, `far from'  means much further away than the contact patch (i.e. $r$ is bigger than a few times $a$).

The qualitative agreement between data and the Cerruti solution means that we can fit our experimental data with the theoretical expressions (equations \ref{eq:bouss_u_Rob}, \ref{eq:bouss_v_Rob}).
To do this, we fit the three parameters $X_0$, $Y_0$ and $F_{lat}/E\equiv \hat{F}$ plus a uniform drift via nonlinear, least-squares fitting (see Materials \& Methods).
In fact, the theory plots in Fig \ref{fig3} show displacement fields using these three fitted parameters -- demonstrating the excellent agreement between theory and data. 
Note, that in the fitting, we exclude data with $r<16 \,\mu$m $\approx 2a$ (see \KK{SI Section S2}). 
This exclusion allows us to ignore parts of the data close to and within the contact area, where the experimental and theoretical data sets should deviate from each other.
Importantly, the fitting is not sensitive to the exact size of the region where we ignore data (see details in \KK{Fig S7}). 

\begin{figure}[tbhp]
    \centering
    \includegraphics[width=1\linewidth]{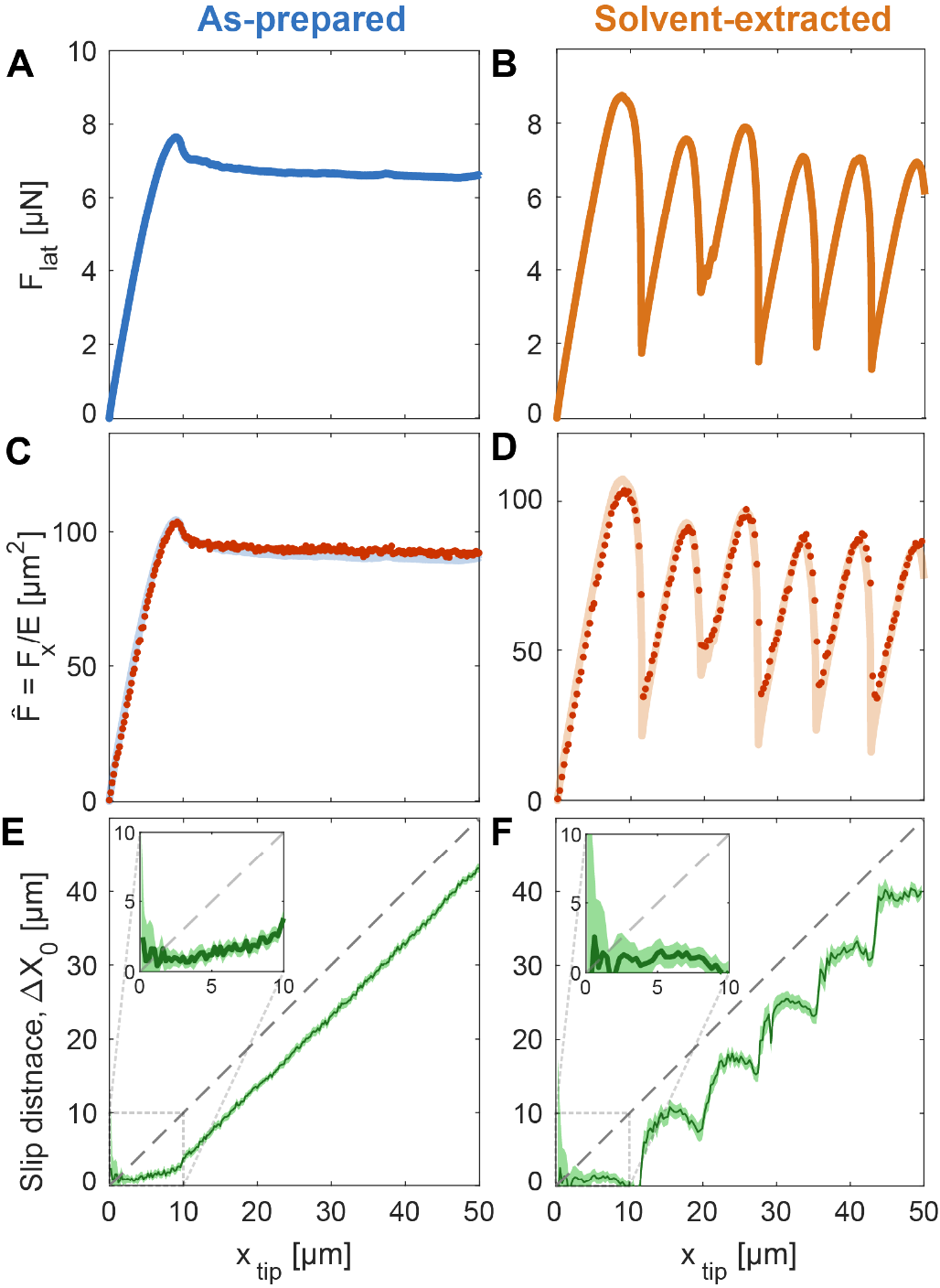}
    \caption{FF-TFM during a friction experiment, for as-prepared (left column) and solvent-extracted (right column) S35:1 silicone. A,B) Lateral force, $F_{lat}$, measured by the AFM. C,D) Fitted values of the scaled force, $\hat{F}=F_{lat}/E$, from continuously fitting the magnitude of substrate displacements during friction experiments (red dots). The curve underneath shows scaled curves $F_{lat}$ from A,B. E,F) Slip distance, $\Delta X_0$, from the same fits. Shaded bands show the uncertainty of the fit, and the dashed line shows the movement of the base of the cantilever. The insets zoom in on the first $10\,\mu$m, before the force maximum in plots A-D.
    }
    \label{fig4}
\end{figure}

While the results in Fig \ref{fig3} show fitting at a single time point, we can also perform the fitting continuously during lateral sliding experiments. 
Figure \ref{fig4}A,B show force traces for such experiments on as-prepared/solvent-extracted silicones, respectively, with a scanning velocity of $\dot{x}_{tip}=0.5\,\mu$m/s.
The traces of $F_{lat}(x_{tip}$) show commonly-observed behaviour for adhesive soft materials, in which the probe initially sticks to the substrate and deforms it elastically ($x_{tip}\lesssim 5\,\mu$m), before starting to slip.
On the as-prepared silicone, upon slipping, the probe continuously slides across the surface -- as demonstrated by $F_{lat}$ reaching a constant plateau (Fig \ref{fig4}A).
In contrast, the probe exhibits strong stick-slip behaviour on the solvent-extracted sample -- as we see from the large peaks and troughs in the force trace (Fig \ref{fig4}B, see \KK{Movie S2} for the full friction loop video).
For both surfaces, we continuously image surface tracer displacements throughout the friction experiments.
This allows us to obtain the fitting parameters $\hat{F}$, and $X_0,Y_0$ as a function of $x_{tip}$.
Figure \ref{fig4}C,D show the fitted values of $\hat{F}$ for the two experiments.
Figure \ref{fig4}E,F show $\Delta X_0=X_0(x_{tip})-X_0(x_{tip}=0)$ -- how much $X_0$ has changed relative to the initial, force-free probe position at the start of the experiment (we show plots of $Y_0$ in the SI \KK{Fig S8}, which is essentially constant).

For both sets of data, there is excellent agreement between the shapes of the measured $F_{lat}$ and the fitted $\hat{F}$.
This is expected, as $\hat{F}$ is simply $F_{lat}/E$.
The high fidelity of the shape agreement both confirms the accuracy of the displacement fitting procedure, and provides a method to extract $E$, not just from a single indentation experiment but from each data point of the friction curve.
We quantitatively match the shapes of $F_{lat}$ and $\hat{F}$ by calculating $\bar{E}=<F_{lat}/\hat{F}>$ (the average is taken over the full length of the friction experiment), and then scaling $F_{lat}$ by this and plotting it with $\hat{F}$ in the Fig \ref{fig4}C,D.
Despite the very different types of frictional behaviour of the two substrates, $F_{lat}/\bar{E}$ (continuous curves) and $\hat{F}$ (dots) match closely, both during the initial sticking (static friction) phase and the following sliding/stick-slip (dynamic friction) phase. 

Crucially, the fitting parameter $\bar{E}$ is a direct measurement of the Young's modulus of the substrates.
For the as-prepared substrate, we obtain $\bar{E} = 73\pm 3$ kPa.
For the solvent-extracted substrate, we obtain $\bar{E} = 80\pm 10$ kPa.
Both these values are in reasonable agreement with values measured via bulk rheology ($E=$71, 96 kPa, respectively).
Indeed, the FF-TFM results show significantly better agreement with rheological data than we found previously using JKR fitting to normal-indentation data (Fig \ref{fig1}C,D).

\begin{figure}
    \centering
    \includegraphics[width=1\linewidth]{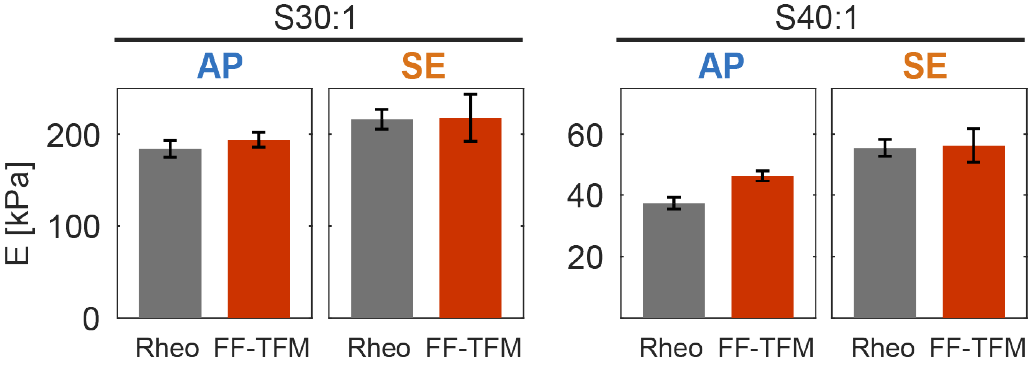}
    \caption{Young's modulus from bulk rheology and from FF-TFM, for stiffer (S30:1) and softer (S40:1) silicones, each solvent-extracted (SE) and as-prepared (AP). Error bars show $\pm5\%$ for rheology, and the standard deviation over the friction loop for FF-TFM.}
    \label{fig5}
\end{figure}

To confirm the accuracy of this technique for measuring stiffnesses, we further show that the FF-TFM approach matches rheological measurements on four additional different substrates.
These substrates are a pair of stiffer silicones (Sylgard 184 with a 30:1 base to curing agent ratio, S30:1) and a pair of softer silicones (Sylgard 184 with a 40:1 base to curing agent ratio, S40:1).
Each pair of substrates consists of an as-prepared silicone, and a solvent-extracted version.
Figure \ref{fig5} shows how, in all cases, there is good agreement between FF-TFM and rheology. 
We also repeated the measurement on all six substrates (as-prepared and solvent-extracted S30:1, S35:1, S40:1) at normal forces ranging from 200-1200 nN. $E$ extracted from FF-TFM does not depend on the normal forces and (see \KK{Fig S9}) and FF-TFM systematically outperforms JKR curve fits (see \KK{SI section S1}).

\subsection{Slippage}
An additional advantage of the FF-TFM technique is that it allows us to precisely detect when and how slippage occurs during friction experiments on soft substrates.
This information is given by the `slip distance' $\Delta X_0$ (Fig \ref{fig4}E,F), representing the lateral distance the probe has slipped across the surface from the force-free reference state.
To understand what $\Delta X_0$ represents, we can imagine stopping a friction measurement and then moving the probe back in the opposite direction until $F_{lat}=0$ (while ensuring no slippage during the return).
If no slippage has occurred, the probe will return to its starting point ($\Delta X_0=0$).
However, if slippage occurs, the probe will not return all the way to its starting point ($\Delta X_0$ will be positive).

We showcase how $\Delta X_0$ reveals stick/slip motion by considering the two data sets in Fig \ref{fig4}E,F.
As described above, the friction experiment on the as-prepared substrate appears to show an initial stick phase ($x_{tip}\lesssim 10\,\mu$m), followed by steady sliding (Fig \ref{fig4}A).
This behaviour is clearly seen in the plot of $\Delta X_0(x_{tip})$ in Fig \ref{fig4}E:
$\Delta X_0$ appears to be fairly constant in the stick phase.
Then, it increases linearly in the sliding phase, following the movement of the probe (dashed line).
Interestingly, however, we see further information upon a closer inspection of $\Delta X_0(x_{tip})$.
In particular, $\Delta X_0$ increases slightly during the initial, `stick' phase (inset, Fig \ref{fig4}E).
This reveals that there is actually a `pre-slip' during this initial phase, which, for such soft contacts, is undetectable from the force trace alone.
On stiffer contacts, by contrast, pre-slip leaves a detectable signature in the force~\cite{Rubinstein2007DynamicsSliding,Maegawa2010PrecursorsLoading,Katano2014NovelSlipping}.
Presumably this pre-slip is promoted by unreacted, free-chain oligomers, which can lubricate the probe/substrate contact.

For the strongly stick-slipping solvent-extracted substrate, $\Delta X_0$ also clearly identifies probe slippage.
Here, $\Delta X_0$ increases in a staircase-like manner.
$\Delta X_0$ jumps up abruptly in value at the same moments where $F_{lat}$ suddenly decreases -- i.e. at points where the probe slips.
$\Delta X_0$ is approximately constant at points where $F_{lat}$ increases smoothly -- indicating sticking.
Interestingly, however, $\Delta X_0$ is not entirely constant during these sticking phases (see \KK{Fig S10} for $\Delta X_0$ evolutions of all samples).
This likely indicates subtle variations in the size and shape of the probe/substrate contact patch and other surface instabilities during stick phases~\cite{Glover2023CreasingFriction,Xu2021GeneralisedDeformation,Lyashenko2024TransitionExperiment}.
Again, these subtle features are undetectable from the force trace alone.

\section{Discussion and Conclusions}

In conclusion, we present a technique, FF-TFM, for extracting the stiffness of soft substrates with micro-scale probes without any of the contact assumptions made by contact models.
These assumptions are hard to avoid.
Even full numerical simulations struggle to reproduce how the substrate deforms, because the conditions at the contact are not known~\cite{Liu2019EffectsSubstrate}.
For millimetre-scale contacts, this can be resolved by imaging displacements inside the contact directly~\cite{Chateauminois2008LocalProbe,Prevost2013ProbingSliding}.
However, this is often not possible for micro-scale contacts.
FF-TFM sidesteps this problem.
It works by performing friction experiments while simultaneously measuring displacements of the free substrate surface away from the contact patch, where the details of the contact no longer matter.

The materials we have characterized here, with $E$ in the range of tens to hundreds of kPa, are widely used in cell biology or tissue engineering but the reported values of their stiffness can vary substantially. 
For example, the same silicone formulation (100:1) has been reported to have $E = 0.1$\,kPa with one indentation method and 41 to 245\,kPa with another one~\cite{Norman2021MeasuringMicroscopy}, which bears critical implications for the interpretation of any data that depends on the value. Part of the variability in the reported data is indeed related to assumptions made by using contact models, where the results may depend on the used probes. If small probes are used, which develop contact radii comparable to the elastocapillary length, then the role of surface tension must be included into the response from indentation~\cite{Style2013SurfaceSolids}. Moreover, the roughness of either surface may change the true contact area and the thus apparent adhesion, affecting the interpretation of the results~\cite{Persson2001TheSolids,Dalvi2019LinkingRoughness}. Our technique is not affeectd by these issues.

Another strong aspect of FF-TFM is its applicability to probes with different geometries, beyond spherical ones, and materials with a broad stiffness range. Provided that the far-field surface displacements are large enough to be detectable within the applicable and measurable lateral forces, then the details of the contact can be ignored and stiffness can be extracted under realistic contacts between irregularly shaped objects and soft substrates, e.g. as in biomedical applications. 
The technique can also be extended to thin samples, 
with $d\lesssim a$. In this case, we can no longer use the simple Cerruti solution for a point force, 
but we need to replace it with the corresponding `Green's function' for point force solutions on finite thickness substrates~\cite{Style2014TractionBiology,Mulligan2018TractionForces}. 
While so far we have applied FF-TFM to transparent materials that can be imaged in transmission, the approach would have to be adapted for opaque materials, where configurations such as those for digital image correlation (DIC) are commonly used to measure strain fields~\cite{Pritchard2013PreciseCorrelation,Chateauminois2008LocalProbe,Liu2025Fatigue-ResistantRobots}.

Beyond measuring stiffness, we have seen that FF-TFM also gives details of probe slippage during friction experiments.
In particular, FF-TFM provides extra information beyond force signals alone, which is of high interest in many cases of applied relevance. When a soft material is joined to a hard one, the interface is where things go wrong. Stresses concentrate there, so the interface often fails prematurely~\cite{Li2025AMechanisms}. This is a central problem in tissue engineering and implant design, where soft tissues or hydrogels have to be attached to bone or to a device. Interestingly, the natural interfaces that survive this (e.g. the meniscus root-bone junction in the knee) make use of slip (among other mechanisms). There, mineral particles slide and roll against the soft collagen matrix and friction dissipates energy, which toughens the interface~\cite{Li2025AMechanisms}. The same point has been made for adhesives, where the amount of time that an interface can slide before it fails changes its resistance to debonding by more than the interfacial energy~\cite{Creton2016FractureReview}. However, this slip is very hard to see, and this has been a long-standing obstacle to understanding soft-hard interfaces~\cite{Creton2016FractureReview}. If the hard component can be glued to a cantilever with a known geometry, and tracers can be placed on the surface of the soft one, our technique offers a way to visualise exactly this, regardless of the probe shape.
We anticipate that this information (e.g. about pre-slip and contact evolution during stick phases) will make FF-TFM a powerful tool for future studies of frictional behaviour for soft contacts, e.g. cells~\cite{Pajic-Lijakovic2024FrictionViscoelasticity}, and tissues~\cite{Li2025ToroidalAnisotropy}.
This is especially true in micro-scale friction, where we are not able to directly image probe substrate contact patches.

Finally, it is worth pointing out two key aspects that motivate further investigations. 
Firstly, FF-TFM assumes a constant stiffness throughout the substrate thickness.
If the substrate has a surface with different properties to the bulk (e.g.~\cite{Bain2026MultiscaleSolids}), then FF-TFM only provides an average stiffness for a layer at the top of the substrate with a thickness of $O(a)$. 
Heterogenous stiffness profiles can thus be probed by performing FF-TFM as a function of normal force, thus probing material properties at different indentation depths. 
Secondly, FF-TFM assumes the substrate material to be linearly elastic. 
For viscoelastic materials, this may break down when the scanning velocity $\dot{x}_{tip}$ is fast enough that viscoelastic effects arise~\cite{Onozuka2024AtomicContact,Xu2020ViscoelasticSurfaces}. 
Furthermore, for solvent-containing materials, such as hydrogels or cartilage, poroelastic solvent flows may occur under the probe during sliding~\cite{Elkington2026AnFriction,Bonnevie2011InArea,Moore2017TribologicalHealth,KalathilAshik2025ThePerspectives}. 
This is expected when $\dot{x}_{tip}\lesssim D/a$, where $D$ is the poroelastic diffusivity. 
In both cases, FF-TFM opens up many exciting possibilities to extract microscale contact stiffness and slippage while measuring friction as a function of different forces and sliding velocities to cover a broad range of relevant conditions without relying on contact models.

\section{Materials and Methods}
\subsection{Substrate preparation}
We machine-mix a two-component silicone kit (Dow Sylgard 184 base and crosslinker) with weight ratios of 30:1, 35:1 and 40:1 to fabricate silicone substrates with different stiffnesses. We pour the mixture in Petri dishes and degas it in a desiccator under vacuum. By varying the volume of the mixture poured in the Petri dish, we control the thickness of the substrate films. The silicone films are cured at 80$^\circ$C overnight and are 2-3 mm thick. We then cut the cured films into 12 squares of approximately 20 mm x 20 mm in size. We transfer 6 pieces on standard glass slides and use them as as-prepared silicone substrates, and solvent-extract the other 6 to remove the unreacted free oligomer chains. 

For the solvent-extraction, we immerse the silicone films in toluene. The films swell and the free oligomers diffuse out of the crosslinked network. To speed up this process, we stir the toluene with an overhead stirrer and refresh it daily over 4 days (see \KK{Fig S11}) using approximately 1 L of toluene in total for 2-3 g of silicone. After discarding the final volume of toluene, we transfer the silicone films to an air/water interface and dry them for 24 hours in a fume hood~\cite{Glover2020ExtractingProperties}. We mount the dried PDMS (now 1-2 mm thick) onto glass slides with the air-facing surface placed face-down so that the water-dried surface is exposed for the measurements. This protocol allows the release of internal stresses during toluene evaporation, which would otherwise cause strong surface wrinkling and irreproducible results for solvent-extracted silicones that are directly affixed to glass slides. We record the sol fractions of all samples by weighing them before and after this solvent-extraction, which were measured as $17.37 \pm 0.03 \%$, $28.24 \pm 0.06 \%$, $33.35 \pm 0.11 \%$ for S30:1, S35:1, and S40:1, respectively.

To image the surface displacements, we deposit fluorescent tracer nanoparticles on half of the films (3 as-prepared, 3 solvent-extracted), without any further surface treatment. The tracers are carboxylate-modified polystyrene nanoparticles (FluoSpheres\texttrademark{}, 100\,nm diameter, abs/em = 505/515\,nm, Thermo Fisher Scientific), diluted from 10\,$\mu$L of stock dispersion into 10\,mL of borate buffer (pH 7.4). We deposit them on the same day the films are ready, i.e. once the as-prepared films have cooled after curing and once the solvent-extracted films have dried. We place a droplet of the dispersion on the silicone surface, let it rest, and remove it by tilting the substrate. The resting time sets the particle density. We use 1 minute on as-prepared and 3 minutes on solvent-extracted silicone, which gives approximately 0.06 tracers per $\mu$m$^2$ in both cases. We then store the films in the dark in a refrigerator until use. 

\subsection{Atomic Force Microscopy}
To perform micro-indentation, we use a spherical probe attached to an AFM cantilever (OMCL-AC160TS, Olympus). The colloidal probe is from a batch of 75-90 $\mu$m soda-lime glass beads (GP0083, Whitehouse Scientific Ltd). We attach the bead to a cantilever in two steps. First, we use an epoxy to temporarily glue the bead to the cantilever, using a micromanipulator mounted on a light microscope (Olympus BX41). Subsequently, we permanently fuse the bead and cantilever by sintering them together in an oven (L3/11/B170, Nabertherm). We heat the oven up to 750$^\circ$C at a ramp rate of 2$^\circ$C/min, hold the temperature for 1 hour, and then leave the oven to cool overnight without opening. After cooling, we clean the whole cantilever in a UV-ozone cleaner (2002A3-US, Ossila) for 30 min. The sintering process slightly deforms the glass bead. To account for this, we measure the shape of the sintered bead by imaging the side of the cantilever via light microscopy. The glass bead retains a spherical shape with an effective diameter of 80 $\mu$m. 

We calibrate the cantilever in two steps. First, before attaching the bead, we obtain the normal spring constant, $k_n$ (in N/m), and the torsional spring constant, $k_\phi$ (in N\,m), from the cantilever dimensions and its resonance in air using the Sader method~\cite{Sader1999CalibrationCantilevers,Green2004NormalCantilevers}. Second, after sintering the bead, we measure the sensitivity of the photodiode to each mode in water after temperature equilibration. For the normal mode, we take force curves against a UV-ozone cleaned glass slide and use the slope of the approach curve, which gives the deflection per volt, $S_n$ (in m/V), giving the normal force conversion factor in N/V ($k_n S_n$). For the torsional mode, we use the following non-contact method~\cite{Mullin2014AMicroscopy}. We record the thermal noise spectrum of the lateral signal and fit the torsional resonance with a simple harmonic oscillator. Together with $k_\phi$ and the torsional arm length (the bead diameter plus half the cantilever thickness, measured to be 75 $\mu$m as imaged by a microscope at an tilted angle), this gives the lateral displacement of the bead per volt (in m/V), and with the torsional stiffness of the lever (in N/m), giving the lateral force conversion factor in N/V. The lateral in-plane bending of the cantilever is also included as a spring in series with the torsion. For the cantilever used here, $k_n=$ 38.6\,N/m and $k_\phi=$ 250\,nN\,m. We use these normal and torsional conversion factors to convert measured V signals to Newton, N. See more details in the \KK{SI, section S1}

All measurements are done on a JPK NanoWizard 3 AFM in fresh MilliQ water. Before each session we clean the colloidal probe in a UV-ozone cleaner (Ossila, L2002A2), calibrate the cantilever, and leave the system to equilibrate in water overnight. We then run friction measurements on the substrate of interest for 1 hour at 0.5\,$\mu$m/s so that the probe surface reaches a steady state before any data are taken.

For indentation measurements, we record force-displacement curves on a $5\times5$ grid of positions spanning 70\,$\mu$m $\times$ 70\,$\mu$m, giving at least 25 curves per condition, with no pause between approach and retraction. The force curves are recorded at a target normal force of 1V (394 nN for as-prepared and 703 nN for solvent-extracted) and a scanning velocity $\dot{z}_{tip}=0.5\,\mu$m/s. The $z$-range is set between 5 and 12\,$\mu$m depending on the substrate stiffness.

For sliding friction experiments, we indent the probe into the substrate at target normal forces ranging from 0.1 to 1\,$\mu$N, let it rest for 1 minute at the target normal force, and then slide the probe over 70\,$\mu$m in each direction (140\,$\mu$m scanned per loop). The lateral force, i.e. the friction force, is measured at a scanning velocity $\dot{x}_{tip}=0.5\,\mu$m/s while the normal force is held constant over the scan. We wait a further minute between loops. This normal force sweep, with its waiting times, is executed using the JPK ExperimentPlanner with a custom script (in the \KK{SI, section S5}). The contact radius, $a$, is estimated from the time lapses described in \KK{SI, section S2 and Fig S12}, by fitting a circle by eye to the region of uniform streaks (maximum displacement) in ImageJ (1.54p), see \KK{Movie S1}.

\subsection{Rheology}
We measure the bulk shear modulus of the same silicone films on a rheometer (MCR 302, Anton Paar) with a 10\,mm parallel plate (PP10, diameter 9.94\,mm), at 25$^\circ$C and under a normal force of 0.1 to 0.5\,N to prevent slip. We record frequency sweeps from 100 down to 0.05\,rad/s at strain amplitudes of 0.2 to 1\%, within the linear regime (\KK{Fig S11}). At the lowest frequencies, which correspond to the strain rates of our sliding experiments ($v/a\sim0.1$\,s$^{-1}$), the loss modulus is less than 10\% of the storage modulus, so the substrates are elastic at the rates used here. We take $G'$ at 0.05\,rad/s, the frequency closest to the strain rate of our experiments, as the reference shear modulus and convert to Young's modulus with $E=3G'$ for an incompressible material, giving 186, 71 and 38\,kPa for as-prepared and 220, 96 and 57\,kPa for solvent-extracted S30:1, S35:1 and S40:1, respectively.

\subsection{Fitting: JKR theory and contact stiffness}
We fit the JKR model to the retraction branch of each force curve, from maximum force ($F_{max}$) to pull-off ($F_{PO}$)~\cite{Ebenstein2011Nano-JKRWater}. The fit has three free parameters: $E$, $W$, $z_0$. The contact point, $z_0$, the probe position where it first touches the surface is left free because it is difficult to locate on soft, adhesive samples. The fit is a nonlinear-least-squares minimisation in MATLAB with multiple starting points and uncertainties are taken from the Jacobian at the solution. More details, including fits for all substrates, are given in the \KK{SI, section S1}.

\subsection{Traction force microscopy during friction measurement: FF-TFM} \label{sec:TFM}
The AFM sits on an inverted microscope (Axio Observer D1, Zeiss) fitted with a spinning-disk confocal unit (CSU-W1, Yokogawa; Prime 95B camera; Oxxius laser). We first locate the silicone surface, then lower the AFM head to find the probe and align its scan path. A bright-field time lapse of the probe over a full loop gives its trajectory in the image, which we use as the initial guess for the force origin $(X_0,Y_0)$. We then record the tracers in confocal mode at 300\,ms exposure while the AFM runs the friction loop. Tracers are tracked with the published MATLAB code track\_RAFT.m (DOI: 10.5281/zenodo.4884065)~\cite{Kim2021MeasuringHysteresis}. The reference frame is the first frame after normal force loading, before any lateral motion. A tracer at $(X_i,Y_i)$ in that frame and at $(x_i,y_i)$ in a later frame has undergone a displacement $(u_{x,i},u_{y,i})=(x_i-X_i,\,y_i-Y_i)$, and the point-force solution is evaluated at the reference positions. For each frame we then find the five parameters $X_0$, $Y_0$, $\hat{F}$, $d_x$ and $d_y$ by minimising

\begin{multline}
  \sum_i \left(u_{x,i} - u_x^C(X_0,Y_0,\hat{F}) - d_x\right)^2 \\
  + \left(u_{y,i} - u_y^C(X_0,Y_0,\hat{F}) - d_y\right)^2,
  \label{eq:bouss_fit}
\end{multline}

\noindent where the uniform offset $(d_x,d_y)$ accounts for drifts of the stage, using a nonlinear least-squares solver in MATLAB. Tracers within a masking radius $R_{ex}$ of the probe are excluded (see \KK{Fig S7}). Uncertainties on $X_0$, $Y_0$ and $\hat{F}$ are taken from the Jacobian at the solution. To obtain $E$, we compare $\hat{F}$ with the friction force measured by the AFM. The two datasets are time synchronised. The point-force solution gives $\hat{F}=F_{lat}/E$, where $F_{lat}$ is the friction force measured by the AFM. From each frame $i$ we thus obtain $E_i=F_{lat,i}/\hat{F}_i$ and $E$ is the mean of $E_i$ over the whole scanning loop.

\section*{Supporting Information Appendix (SI)}
The SI Appendix contains details of the JKR analysis, including a comparison of the conventional curve fit with the two-point method (nano-JKR) and a discussion of technical sources of error in indentation measurement by AFM (section S1); a comparison of contact radii predicted by JKR theory with those estimated from tracer images (section S2); FF-TFM fitting of all six samples, including the dependence of $E$ on the exclusion radius and normal force, and slip distances (section S3); practical notes on solvent extraction of silicone films (section S4); the AFM measurement protocol/script (section S5); rheology data; and legends for Movies S1 and S2. All data will be made publicly available.

\subsection*{SI Datasets}
\textbf{Datasets S1 and S2.} Indentation curves on S35:1 AP and SE at 0.5 $\mu$m/s.
\textbf{Datasets S3 to S5.} Friction data, imaging parameters with frame times, and tracer trajectories of the FF-TFM experiment on S35:1 AP.
\textbf{Datasets S6 to S8.} As Datasets S3 to S5, for S35:1 SE.

\subsection*{SI Movies}
\textbf{Movie S1.} Confocal time lapse of the tracers on solvent-extracted S35:1 during the friction experiment of Fig \ref{fig2}B,C.
\textbf{Movie S2.} Confocal time lapse of the tracers on solvent-extracted S35:1 over a full friction loop, showing the stick-slip motion of the contact.

\begin{acknowledgments}
Authors thank Dr. Kirill Feldman for supporting bulk rheometry measurements, Dr. Nicholas Spencer, Dr. Robert Carpick, Dr. Antoine Sanner, Dr. Simon Scherrer, Dr. Jan Van Dokkum, Lukas Woolley, Tizian Keller for fruitful discussions.
\end{acknowledgments}

%

\end{document}